\documentclass[aps,preprint,onecolumn,showpacs,amsmath,amssymb]{revtex4}
\usepackage{graphics}

\newcommand{\ie}{\textit{i. e.~}}
\begin{document}

\title{Decoherence induced by an interacting spin environment in the transition from integrability to chaos}

\author{A. Rela\~no}
\email{armando@iem.cfmac.csic.es} \affiliation{Instituto de Estructura de la Materia, CSIC, Serrano 123, E-28006
Madrid, Spain}

\author{J. Dukelsky}
\email{dukelsky@iem.cfmac.csic.es} \affiliation{Instituto de Estructura de la Materia, CSIC, Serrano 123, E-28006
Madrid, Spain}

\author{R. A. Molina}
\email{molina@iem.cfmac.csic.es} \affiliation{Instituto de Estructura de la Materia, CSIC, Serrano 123, E-28006
Madrid, Spain}

\date{\today}
\begin{abstract}
We investigate the decoherence properties of a central system composed of two spins 1/2 in contact with a spin
bath. The dynamical regime of the bath ranges from a fully integrable integrable limit to complete chaoticity. We
show that the dynamical regime of the bath determines the efficiency of the decoherence process. For perturbative
regimes, the integrable limit provides stronger decoherence, while in the strong coupling regime the chaotic limit
becomes more efficient. We also show that the decoherence time behaves in a similar way. On the contrary, the rate
of decay of magnitudes like linear entropy or fidelity does not depend on the dynamical regime of the bath. We
interpret the latter results as due to a comparable complexity of the Hamiltonian for both the integrable and the
fully chaotic limits.
\end{abstract}
 \maketitle

\section{Introduction}

Real quantum systems always interact with their environment. This interaction entails that the system, initially
in a pure state, becomes entangled with the environment and decays into an incoherent mixture of several states.
This phenomenon, called decoherence, is an essential feature of quantum mechanical systems.

From a fundamental point of view, decoherence provides a theoretical basis for the quantum-classical transition
\cite{Zurek:03}, emerging as a possible explanation of the quantum origin of the classical world. From a practical
point of view, it is a major obstacle for building a quantum computer \cite{Nielsen} since it can produce the loss
of the quantum character of the computer. Therefore, a complete characterization of the decoherence process and
its relation with the physical properties of the system and the environment, such as the strength of the
system-bath interaction, characteristic times of the bath, or the presence of quantum phase transitions or quantum
chaos, is needed for both fundamental and practical purposes.

Connections between decoherence and quantum chaos have been previously studied. However, a universal theory has
not yet been found. One line of argument establishes a link between the decoherence process and the Loschmidt echo
\cite{Cucchietti:03}, claiming that for a quantum system with a classically chaotic Hamiltonian the rate at which
the environment degrades information on the initial state becomes independent of the system-environment coupling
strength \cite{lyapunov}. Another point of view, in some way contrary to the former, but generally accepted,
states that a chaotic bath leads to faster and stronger decoherence than an integrable one
\cite{Zurek:01,Dobrovitski:03,Blume:03}. One significative manifestation of this phenomenon is the dependence of
the decoherence time, \ie the time for which the initial correlations in the central system are lost due to
decoherence, with the system-bath coupling strength $\lambda$. Some authors have found that for regular baths
decoherence rate is proportional to $\lambda^2$, while chaotic or unstable ones display a considerable  weaker
dependence on $\lambda$ \cite{Dobrovitski:03,Blume:03}. A numerical study over a quantum walker with a complex
coin has shown that, though a chaotic and a regular environments may not be distinguishable in the short-time
evolution, the chaotic one continues to be effective over exponentially longer time scales, whereas the regular
bath saturates much sooner \cite{Paz:06}. A similar study on the Dicke model at weak coupling shows that the
entanglement is smaller if the system is initially in a regular orbit than if it is in an irregular one
\cite{Hou:04}. However, exceptions for this general behavior are well known \cite{regular}. It is also argued
that, when the system-bath interaction becomes extremely small, so that the perturbation theory is applicable, the
regular bath leads to a faster decoherence than the chaotic one \cite{regular_decoherencia}. Numerical studies of
many-spin system show that a chaotic bath generates stronger and faster decoherence than an integrable one for
strong enough coupling. However, the result is opposite in the perturbative regime \cite{Lages:05}.

In this paper we study the connection between decoherence and quantum chaos in a many-body spin systems. We follow
the methodology proposed in \cite{Lages:05} in order to test if the conclusions obtained there are applicable to a
broader class of spin systems and, thus, can be postulated as generic. We use a Hamiltonian for the bath that
depends on many arbitrary, real and independent parameters, and whose dynamical regime is independent of the
specific values of these parameters. The integrable limit is defined as a random realization of the XYZ Gaudin
magnet \cite{Gould:02}, characterized by the existence of as many integrals of motion as quantum degrees of
freedom. The transition to a chaotic regime is modelled by a single control parameter interpolating between the
integrable Hamiltonian and a fully chaotic one. In both limits, and along the whole transition, the complexity of
the Hamiltonian, understood as the number of different relevant terms, remains comparable, contrary to most of the
previously studied systems, for which the regular limit is represented by a simplified Hamiltonian. For example,
in \cite{Paz:06} the integrable limit is characterized by an independent evolution of each spin of the bath, and
in \cite{Lages:05} it is reached when the Hamiltonian of the bath reduces to a site-dependent magnetic field, with
a negligible interaction between different spins.

In this work, we show that, for a wide class of spin Hamiltonians, the integrable limit generates decoherence
more efficiently if the system-bath coupling strength is small, while the chaotic limit becomes more efficient
when the coupling is larger. We also show that this conclusion can be extended to the decoherence time.
Nevertheless, the transition from integrability to chaos in terms of magnitudes related to decoherence is not so
smooth as it is on spectral statistics. Moreover, the rate of decay of the fidelity and the linear entropy does
not depend on the dynamical regime of the bath, contrary to what it is usually claimed.

The paper is organized as follows. In Sec. II, we describe the model and analyze its dynamical regime by means of
spectral statistics. In Sec. III, we study the efficiency of the decoherence process, illustrating  the connection
between the dynamical regime of the bath and some characteristic measures of decoherence, like the non-diagonal
elements of the system reduced density matrix and the linear entropy. In Sec. IV we study a quantitative
characterization of the chaoticity of the bath in a perturbative regime using the linear entropy and the Loschmidt
echo.  We also study the relation between the onset of chaos and the decoherence time. Finally, in Sec. V we
summarize our results.

\section{The model}

We will consider a central system composed by two interacting spins 1/2, {\bf S$_1$} and {\bf S$_2$}, and a bath
composed of a large number of 1/2 spins {\bf I$_k$} \cite{nota1}.  The central system and the bath evolve with the
following Hamiltonian:
\begin{equation}
H = H_S + H_{SB} + H_B,
\end{equation}
where $H_S$ is the self-Hamiltonian of the system, $H_{SB}$ the interaction between the system and the bath, and
$H_B$ the Hamiltonian of the bath. For $H_S$ and $H_{SB}$ we use the Hamiltonians
\begin{equation}
H_S = J \mathbf{S}_1 \cdot \mathbf{S}_2,
\end{equation}
and
\begin{equation}
H_{SB} = \mathbf{S}_1 \cdot \sum_k a_k \mathbf{I}_k.
\end{equation}
The interaction between the central system and the bath is carried out by a single spin of the system
$\mathbf{S}_1$; the other spin, $\mathbf{S}_2$, is affected by the bath indirectly, through its interaction with
$\mathbf{S}_1$, governed by $H_S$. These kind of models are useful to describe, for example, the destruction of
Kondo effect by decoherence \cite{Katsnelson:03}.

For the bath Hamiltonian $H_B$ we use an XYZ model with long range interactions
\begin{equation}
H_B = \sum_j \epsilon_j H_j,
\end{equation}
where
\begin{equation}
H_j = \sum_{k = 1 \ne j}^N A_{jk} I^x_j I^x_k + B_{jk} I^y_j I^y_k + C_{jk} I^z_j I^z_j, \label{eq:HB_general}
\end{equation}
and $\{ \epsilon_j \}$ are free parameters.

With this generic Hamiltonian we will describe a complete transition from integrability to chaos, depending on the
properties of matrices $\mathbf{A}$, $\mathbf{B}$ and $\mathbf{C}$. The integrable limit is obtained when the
Hamiltonians (\ref{eq:HB_general}) fulfilled the conditions of the XYZ Gaudin integrable model. In this limit the
$N \times N$ matrices $\mathbf{A}$, $\mathbf{B}$ y $\mathbf{C}$ are defined in terms set of $N$ arbitrary
parameters $\{ z_j \}$, according to the following identities:
\begin{eqnarray}
A_{jk} &= \dfrac{1 + \kappa \; \text{sn}^2 (z_j - z_k)}{\text{sn} (z_j - z_k)},  \nonumber \\
B_{jk} &= \dfrac{1 - \kappa \; \text{sn}^2 (z_j - z_k)}{\text{sn} (z_j - z_k)}, \label{xyz} \\
C_{jk} &= \dfrac{\text{cn} (z_j - z_k) \; \text{dn} (z_j - z_k)}{\text{sn} (z_j - z_k)}, \nonumber
\end{eqnarray}
where $\text{sn} (u) \equiv \text{sn} (u,\kappa)$ is the Jacobi elliptic function of modulus $\kappa$, $0 \leq
\kappa \leq 1$, and $\text{cn}(u)$ and $\text{dn} (u)$ are related to $\text{sn}(u)$ by $d \; \text{sn}(u) / du =
\text{cn}(u) \; \text{dn} (u)$. The XYZ Gaudin model can be solved exactly by Bethe ansatz \cite{Gould:02}. There
as many independent Hamiltonians (\ref{eq:HB_general}) as quantum degrees of freedom and, with the definition
(\ref{xyz}) they commute among themselves, $[H_i, H_j] \: \forall i,j = 1, \ldots, N$ for arbitrary values of the
parameters  $\{ z_j \}$. Therefore, they constitute a complete set of integrals of motion \cite{nota2}.

The transition from integrability to the fully chaotic limit is performed by a single-parametric perturbation of
the matrices defined above. If the amplitude of the perturbation is small, the resulting Hamiltonian is close to
integrability; for increasing values of the parameter, the Hamiltonian approaches the fully chaotic limit. Such a
perturbation can be achieved with the following identities
\begin{eqnarray}
A'_{jk} &=&(\cos \alpha) \; A_{jk} + (\sin \alpha) \; R^1_{jk} \nonumber,  \\
B'_{jk} &=& (\cos \alpha) \; B_{jk} + (\sin \alpha) \; R^2_{jk}, \label{transicion} \\
C'_{jk} &=& (\cos \alpha) \; C_{jk} + (\sin \alpha) \; R^3_{jk}, \nonumber
\end{eqnarray}
where $R^1_{jk}$, $R^2_{jk}$ and $R^3_{jk}$ are random antisymmetric matrices, and $0 \leq \alpha \leq \pi/2$.
Therefore, for $\alpha>0$ the algebraic structure of the integrable system is lost, in a similar way as the
geometric structure of a classical integrable system is broken when it is perturbed (see \cite{Relano:04} for a
complete discussion about this definition of quantum integrability and its connection with spectral statistics).
Note, however, that $H_B$ conserves its complexity, remaining a truly XYZ model along the whole transition; none
of the X, Y and Z terms becomes negligible in the integrable limit. In consequence, this model allows to study the
influence of the dynamical regime of the bath in decoherence process independently of its complexity.

\subsection{Spectral statistics of the bath}

The concept of quantum chaos still lacks a clear definition. Usually, a quantum system is said to be regular or
chaotic depending on the statistical properties of its spectrum. Using a semiclassical approximation, Berry and
Tabor showed than the statistical properties of the spectrum of a generic quantum integrable system are well
described by an uncorrelated Poisson distribution \cite{Berry:77}. On the other hand, Bohigas {\it et al.}
conjectured that the statistical properties of the spectrum of a generic quantum chaotic system coincide with
those of Random Matrix Theory \cite{Bohigas:84}. Therefore, the statistical properties of the spectrum of a
quantum system are considered as a main signature of chaos in quantum mechanics (for a recent review see
\cite{Haake}).

For a general quantum system, the level density $\rho(E)$ can be separated into a smooth part $\overline{\rho(E)}$
and a fluctuating part $\widetilde{\rho(E)}$. The former depends on the specific properties of the Hamiltonian,
while the latter is universal depending only on the dynamical regime of the system \cite{Haake}. Therefore, in
order to determine whether a quantum system is regular or chaotic from the statistical properties of its spectrum,
it is necessary to extract the fluctuating part of the density. This is due by means of a procedure called {\it
unfolding}, which maps every energy level $E_i$ to a dimensionless magnitude $\zeta_i$,
\begin{equation}
\zeta_i = \overline{N(E_i)},
\end{equation}
where $N(E)$ is the accumulated level density
\begin{equation}
N(E) = \int_{-\infty}^{E} d x \rho (x).
\end{equation}
This map can be done analytically in a few simple systems, like quantum billiards or Random Matrix Ensembles, but
in general it is a difficult task. In this paper, we have performed the unfolding by approximating
$\overline{N(E)}$ with a set of Chebyshev polynomials by means of a least-squares fit.

The most simple and widely used spectral statistic is the nearest-neighbor spacing distribution $P(s)$, {\it i.e.}
the probability distribution of the nearest-neighbor spacing sequence $s_i = \zeta_{i+1}- \zeta_{i}$. For a
regular quantum system the distribution follows a Poisson $P(s) = \exp (-s)$, while for a quantum chaotic system
it follows a Wigner distribution $P(s) = \left( \pi / 2 \right) s \exp \left (- \pi s^2 / 4 \right)$. Note that in
both cases $\left< s \right>=1$.

\begin{figure}[!]
\begin{center}
\rotatebox{-90}{\scalebox{0.28}[0.3]{\includegraphics{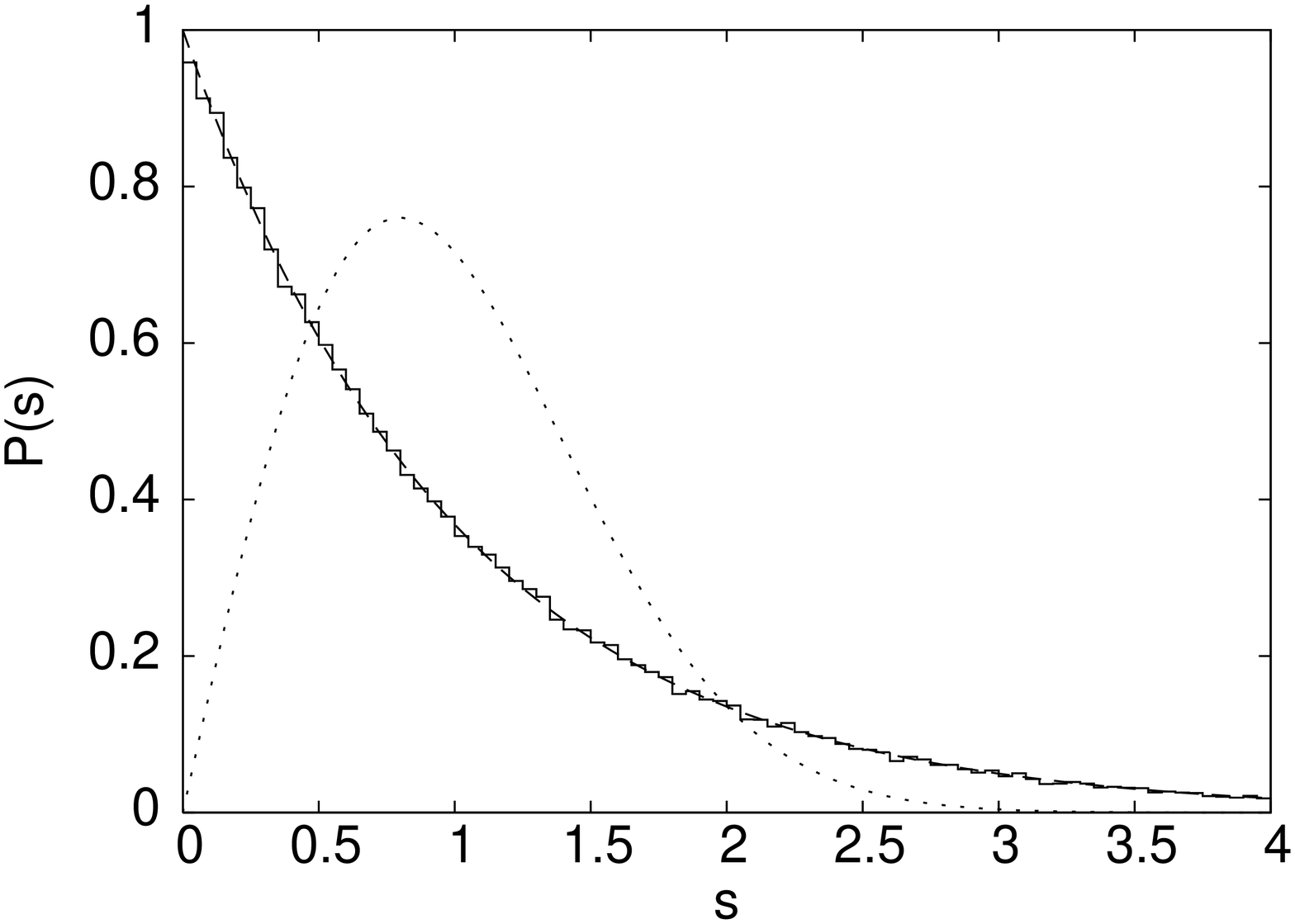}}} \\
\rotatebox{-90}{\scalebox{0.28}[0.3]{\includegraphics{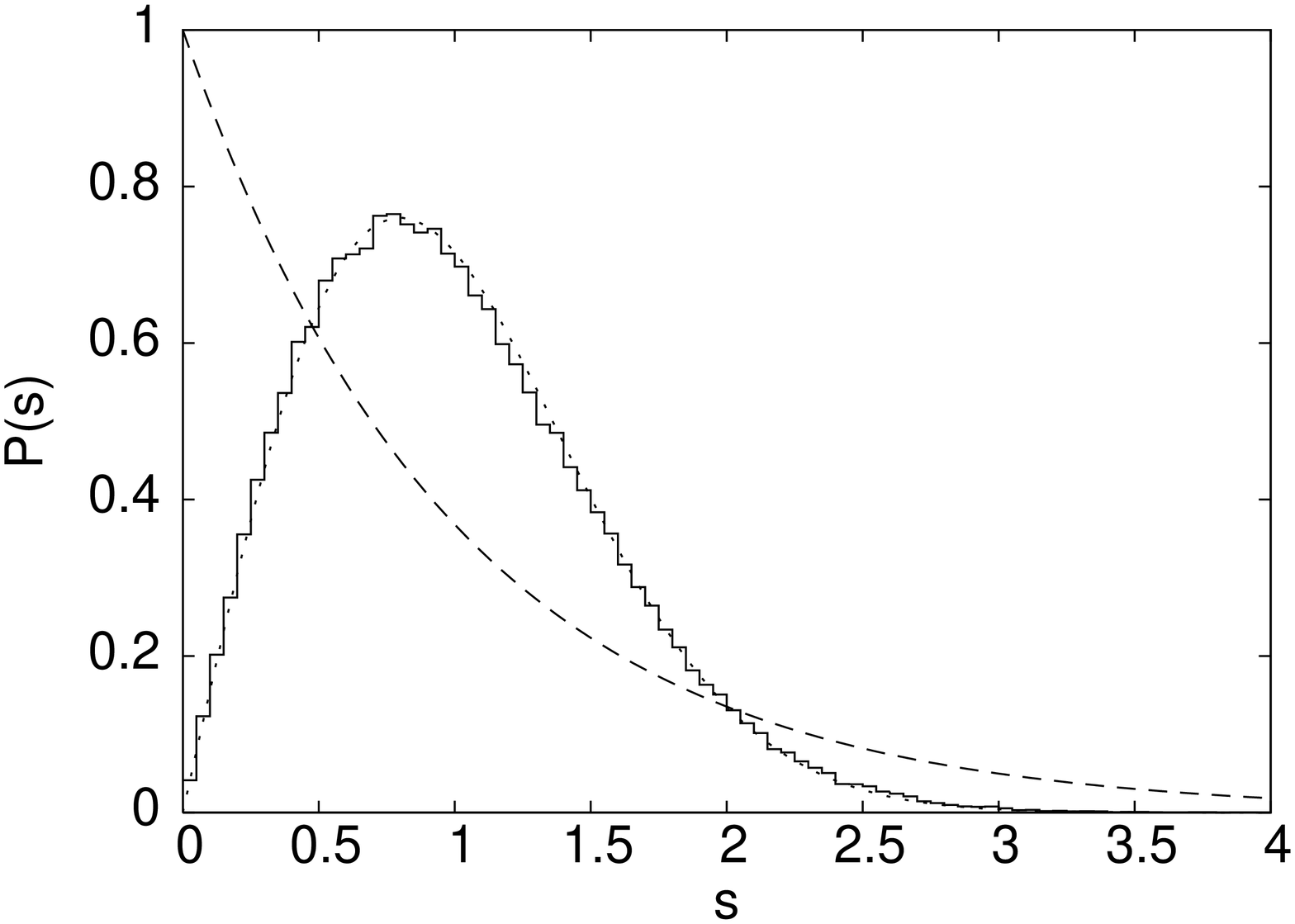}}} \\
\caption[]{P(s) distributions for regular $\alpha=0$ (upper panel),  and chaotic $\alpha=\pi/2$ (lower panel) limits of $H_B$. In both cases
there are $N=13$ spins, and the histogram is built by collecting 50 different cases, characterized by different
sets of parameters $\{ \epsilon_j \}$.  The dashed line corresponds to the Poisson distribution, while the dotted
line represents the Wigner distribution.} \label{fig:pds}
\end{center}
\end{figure}

In Fig. \ref{fig:pds} we show the $P(s)$ distribution for a set of 50 different realizations of the regular
($\alpha=0$) and fully chaotic ($\alpha=\pi/2$) limits of bath Hamiltonian $H_B$ with $N=13$ spins. The random
matrices $\mathbf{R^1}$, $\mathbf{R^2}$ and $\mathbf{R^3}$ are defined in terms of Gaussian random variables with
zero mean, and variance equal to
\begin{equation}
\sigma(\mathbf{R^1}) = \sigma (\mathbf{R^2}) = \sigma (\mathbf{R^3}) = \frac{\sigma(\mathbf{A}) +
\sigma(\mathbf{B}) + \sigma(\mathbf{C})}{3}.
\end{equation}
Each realization is obtained by choosing an independent set of $\{ \epsilon_j \}$ parameters, by means of Gaussian
random variables with zero mean and $\sigma=1$. For all realizations the set of $\{ z_j \}$ parameters is fixed to
$z_j = 3.71 \sqrt{ j/N}$, in order to cover the whole period of the Jacobi elliptic functions, and the modulus of
the Jacobi elliptic functions is fixed to $\kappa = 0.5$.  As can be seen in Fig. \ref{fig:pds} the regular limit
clearly follows the Poisson distribution, while the chaotic limit is perfectly described by the Wigner
distribution.

\begin{figure}[!]
\begin{center}
\rotatebox{-90}{\scalebox{0.28}[0.3]{\includegraphics{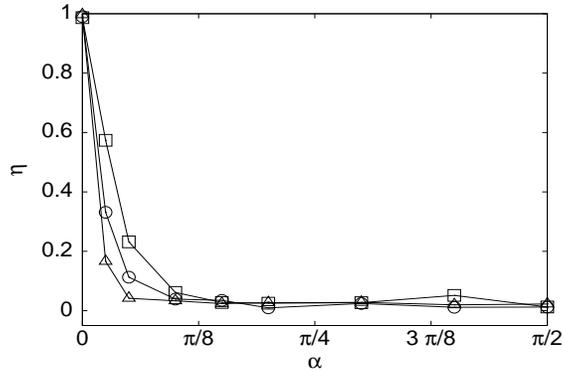}}} \\
\caption[]{Parameter $\eta$ in function of $\alpha$ for $N=9$ (squares), $N=11$ (circles) and $N=13$ (triangles).}
\label{fig:eta}
\end{center}
\end{figure}

In order to quantify the degree of chaoticity of the bath as function of the parameter $\alpha$ it is useful to
calculate the following quantity
\begin{equation}
\eta = \frac{ {\displaystyle \int}_0^{s_0} ds \; \left( P(s) - P_{Wigner} (s) \right)}{ {\displaystyle
\int}_0^{s_0} ds \; \left( P_{Poisson}(s) - P_{Wigner}(s) \right)},
\end{equation}
where $s_0=0.472913$ determines the first intersection of Poisson and Wigner distributions. This parameter
transits from $\eta=1$ to $\eta=0$ when the system moves from integrability to chaos. Therefore, the curve
$\eta(\alpha)$ shows how fast or slow is this transition. In Fig. \ref{fig:eta} we show the value of $\eta$ as a
function of $\alpha$ for three different sizes of the bath, $N=9$, $N=11$ and $N=13$. For the three cases the
system very fast approaches to chaos for small values of $\alpha$. The transition is increasingly faster for the
larger bath sizes.

The $P(s)$ distribution describes short-range correlations, since it measures the fluctuations in distances
between consecutive levels. To properly determine the chaoticity of a quantum system, it is also necessary to
study the long-range spectral correlations. There are several statistics to measure this long-range correlations.
The most commonly used are $\Sigma_2 (L)$ and $\Delta_3 (L)$ statistics \cite{Haake}. In this paper, we will use
instead the $\delta_n$ statistic, defined from the unfolded energies as \cite{delta_n}
\begin{equation}
\delta_n = \zeta_{n+1} - \zeta_1 -n.
\end{equation}
This statistic measures the fluctuations of the unfolded energy levels $\{ \zeta_i \}$ from their average
value. In particular, we are interested in its power spectrum
\begin{equation}
P^{\delta}_k = \frac{1}{N} \left| \sum_{n=1}^N \delta_n \exp (-2 \pi i n k/N) \right|^2,
\end{equation}
which is proportional to $1/k^2$ for regular systems, and $1/k$ for chaotic systems \cite{delta_n}. This statistic
is simple to compute, and it is more sensitive to the dynamical regime of the system than the $P(s)$ statistic
(see \cite{Gomez:05} for a detailed discussion of this point).

\begin{figure}[!]
\begin{center}
\begin{tabular}{cc}
\rotatebox{-90}{\scalebox{0.28}[0.3]{\includegraphics{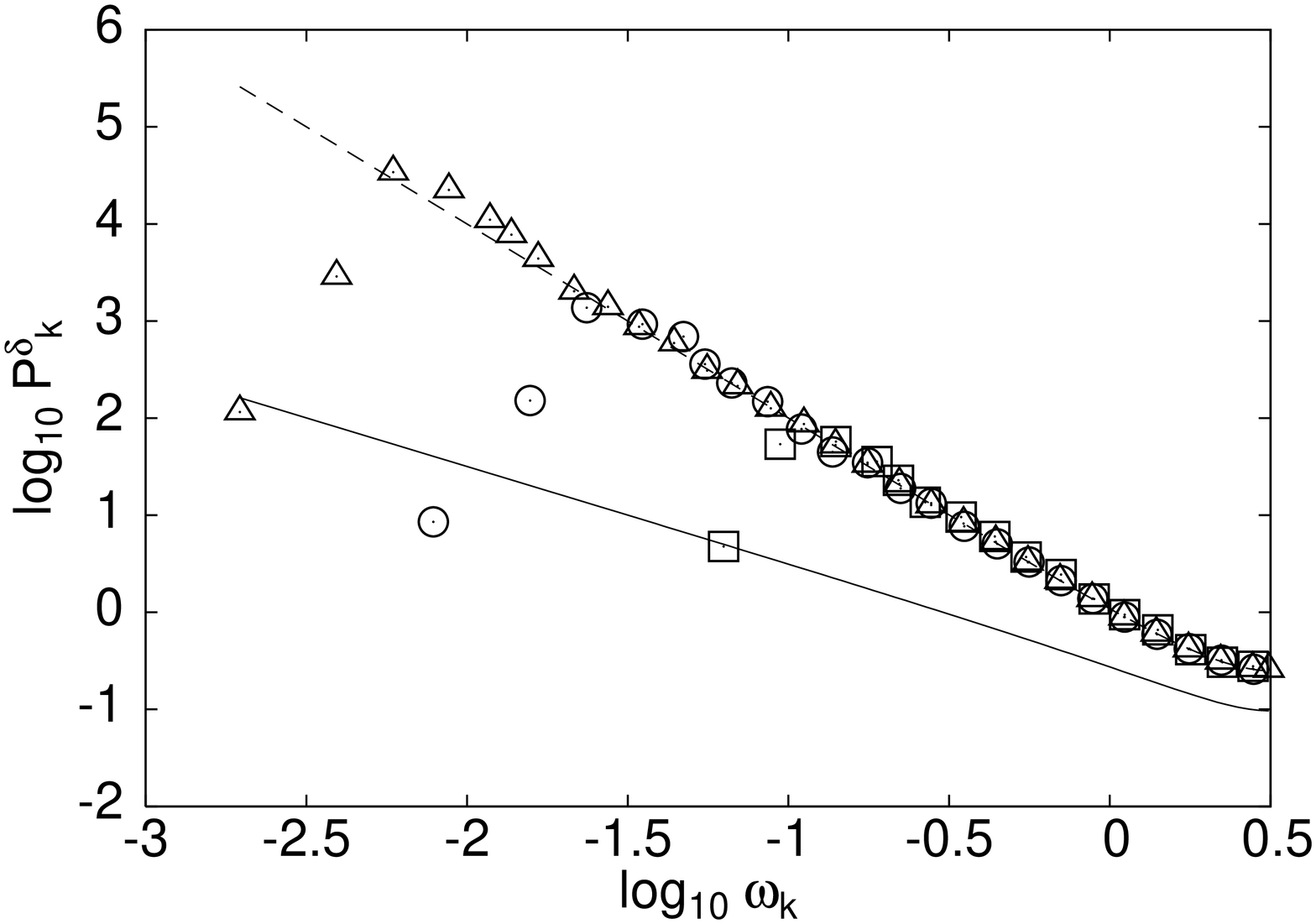}}} & \rotatebox{-90}{\scalebox{0.28}[0.3]{\includegraphics{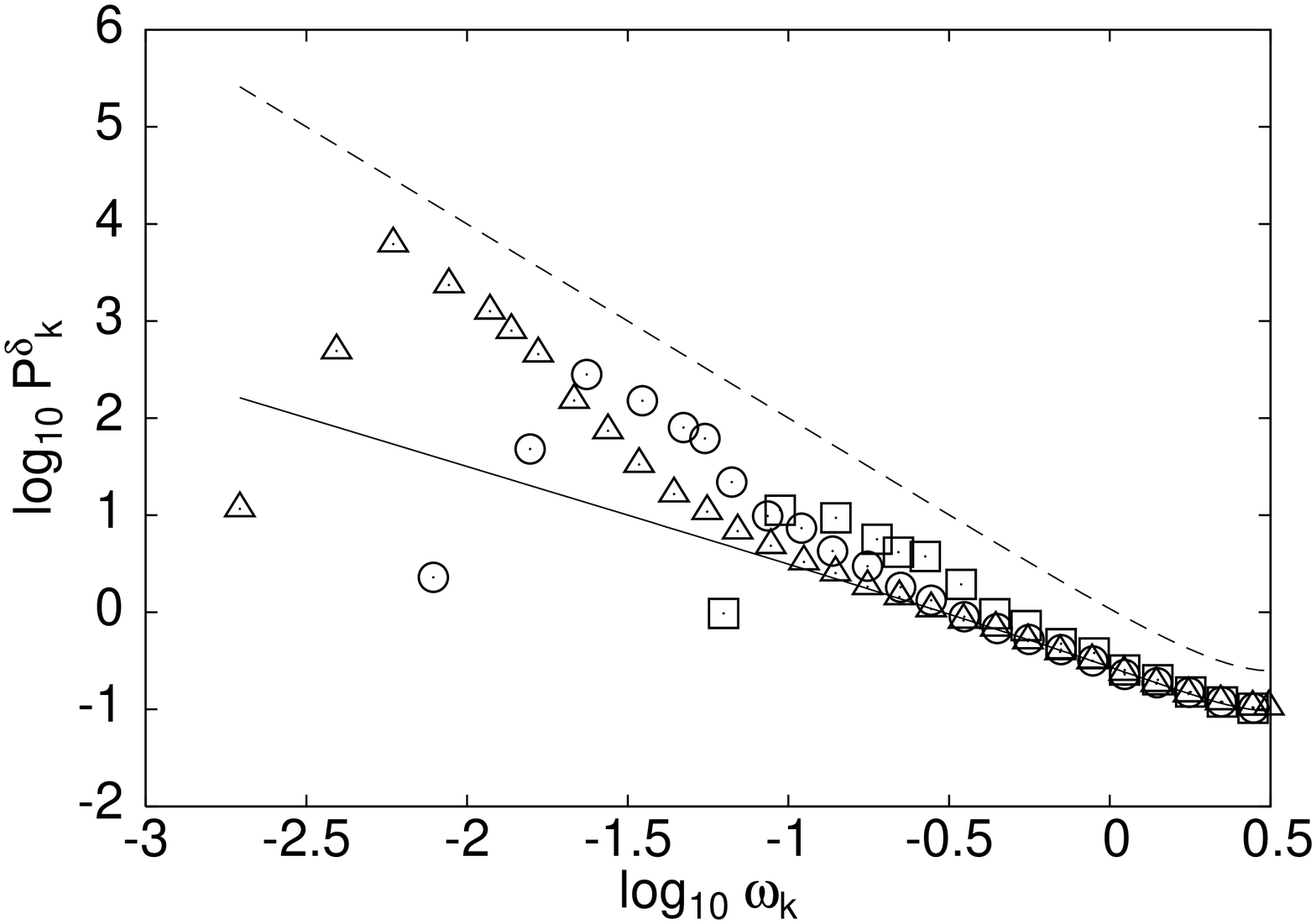}}} \\
\rotatebox{-90}{\scalebox{0.28}[0.3]{\includegraphics{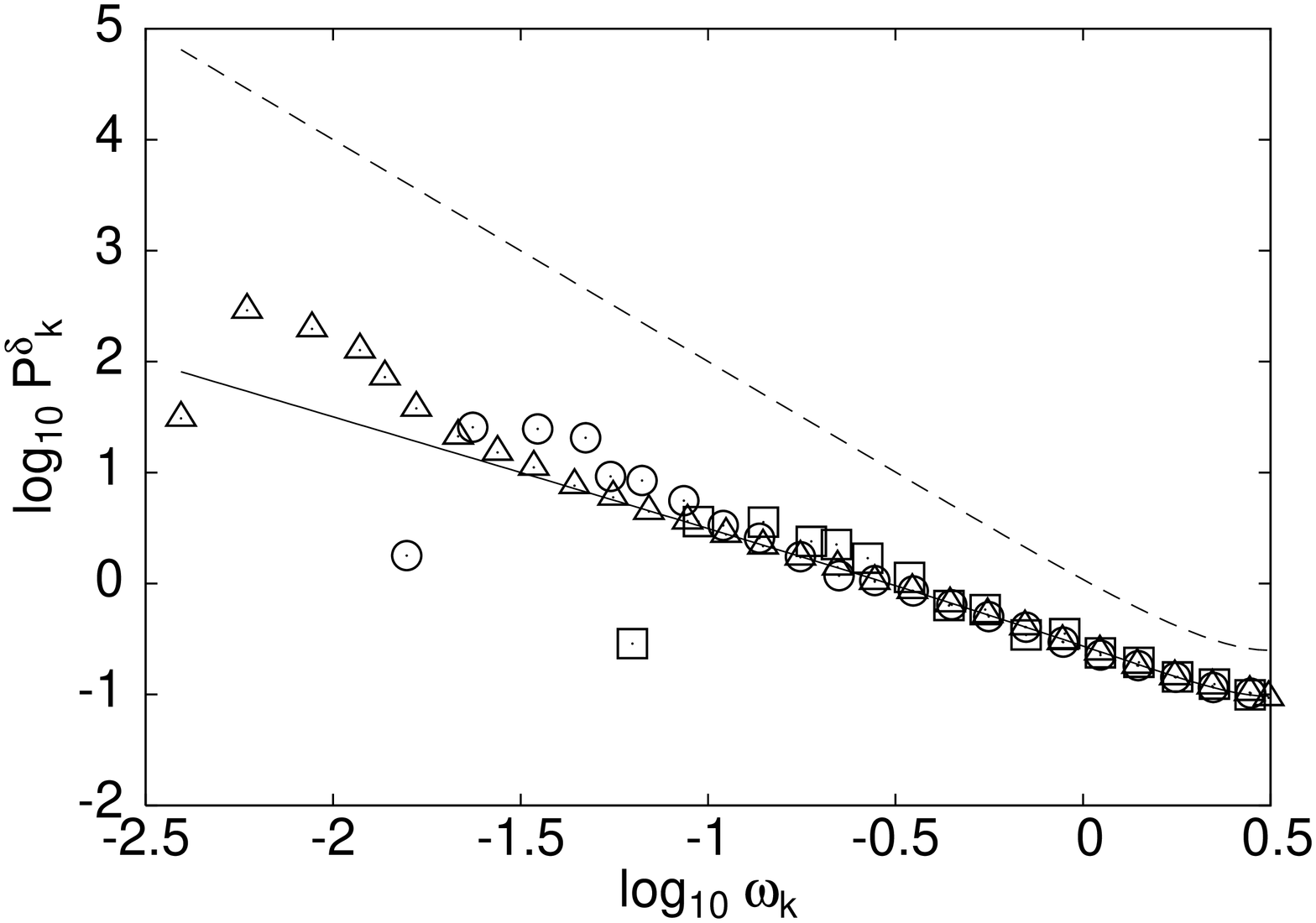}}} & \rotatebox{-90}{\scalebox{0.28}[0.3]{\includegraphics{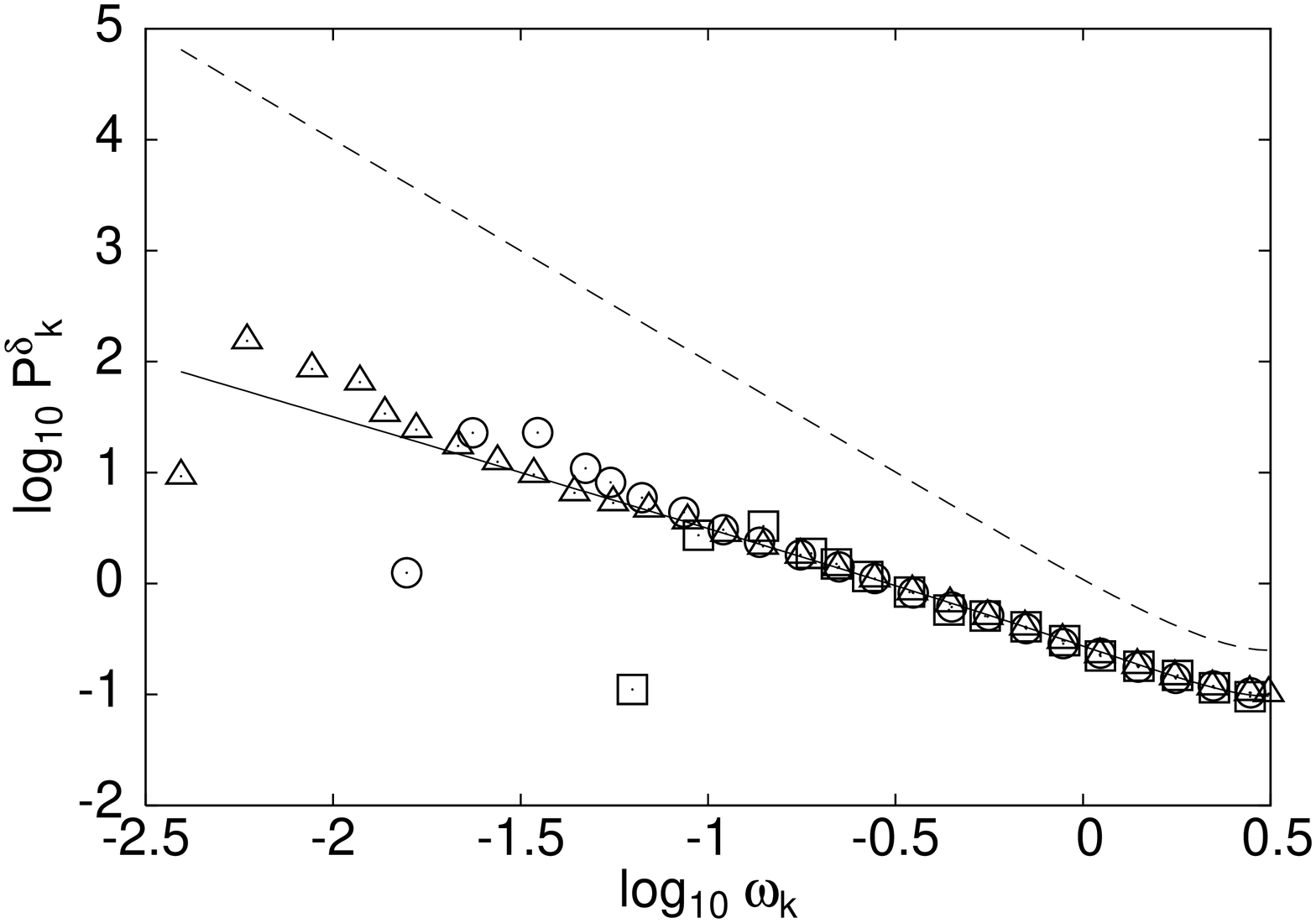}}} \\
\end{tabular}
\caption[]{Power spectrum of $\delta_n$ statistic for $\alpha=0$ (upper left panel), $\alpha=\pi/10$ (upper right panel), $\alpha=\pi / 5$ (lower left panel), and $\alpha=\pi/2$ (lower right panel). Squares correspond to $N=9$;
circles correspond to $N=11$; and triangles correspond  to $N=13$. The theoretical value for integrable system is plotted with a dashed line, and the theoretical value for chaotic systems, with a solid line.} 
\label{fig:delta_n}
\end{center}
\end{figure}

We show in Fig. \ref{fig:delta_n} the power spectrum of $\delta_n$ statistic for $\alpha=0$, $\alpha=\pi/10$,
$\alpha=\pi /5$ and $\alpha=\pi / 2$ in a double logarithmic scale.  The x axis is plotted in function of
$\omega_k = 2 \pi k / N$, which ranges from $\omega=0$ to $\omega=\pi$ independently of the size of the spectrum,
and therefore allows to look for finite size effects quite easily. Both, the regular and the chaotic limit,
 closely follow the theoretical lines, except in the low frequency region.  This region is spoiled by the unfolding
procedure (see, for example, \cite{Gomez:02} for a complete discussion about the misleading effects due to the
unfolding). Note also than the smallest frequency available is $\omega_1 = 2 \pi / N$, and thus larger systems
cover a wider range of frequencies. For intermediate values of $\alpha$, the result is similar to the obtained
with the $\eta$ parameter, {\it i.e.} the transition to chaos is fast, and for a fixed value of $\alpha$ larger
sizes are more chaotic. However, it is also seen that the transition for the power spectrum of $\delta_n$
statistic appears to be slower than the description given in terms of the parameter $\eta$. In Fig. \ref{fig:eta},
the parameter $\eta$ identifies the system as almost chaotic for $\alpha=\pi/10$ and $N=13$, whereas in Fig.
\ref{fig:delta_n} (upper right panel) it is clearly seen that the $P^{\delta}_k$ statistic is still far from the
chaotic limit (lower right panel). These differences can also be seen for $\alpha=\pi/5$, but they might be due to the
spurious effects of the unfolding procedure.

In conclusion, the bath Hamiltonian as defined in the previous subsection develops a complete transition from
integrability to chaos, manifested in both short-range and long-range spectral statistics. Moreover, this
transition is smooth and monotonous with the parameter $\alpha$.  The transition is faster for larger systems
sizes and therefore a normalization $\alpha(N)$ is required for the results to be independent of the size of the
bath. Hence, $H_B$ is a good candidate to study the connection between decoherence and quantum chaos in finite
size spin bathes. In the thermodynamic limit, the transition from integrability to chaos may be sharp for $\alpha
\gtrsim 0$.

\subsection{Density of states of the bath}

\begin{figure}[!]
\begin{center}
\rotatebox{-90}{\scalebox{0.28}[0.3]{\includegraphics{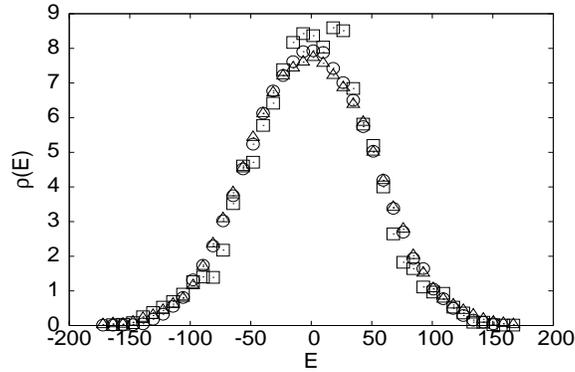}}} \\
\caption[]{Density of states of $H_B$ with $\alpha=0$ (squares), $\alpha=\pi / 10$ (circles) and $\alpha=\pi/2$ (triangles).}
\label{fig:densidades}
\end{center}
\end{figure}

Prior to the analysis of the connection between chaos and decoherence using the bath Hamiltonian defined above, it
is important to check that the parameter $\alpha$ modifies the chaotic properties of the system without altering
in a significant way the density of states of the bath. Changes in the density of states may have an important
influence on decoherence processes \cite{Marquardt:06}. In Fig. \ref{fig:densidades} we show the density of states
of $H_B$ for three different values of $\alpha$, corresponding to the regular limit, the fully chaotic limit and
an intermediate case. For these calculations, the set of parameters $\{ \epsilon_j \}$ has not been chosen
randomly, but according to $\epsilon_j = \cos \left( \sqrt{2} j \right)$. It can be seen that there are no
important differences between the three cases under study. Similar results are obtained for other values of
$\alpha$ (not shown). Therefore, we can conclude that $\alpha$ determines the degree of chaoticity of the spin
bath, without altering significantly the density of states of the spin bath.  However, it is important to note
that the density of states of the bath is quite sensitive to the parameters of the model Hamiltonian $H_B$,
specially to the set $\{ z_j \}$. The use of these kind of Hamiltonians to study the transition from integrability
to chaos requires a careful definition of the parameters.

\section{Decoherence and Chaos: long time evolution and decoherence efficiency}

\subsection{General description of time evolution}

We will study the time evolution of the system from an initial product state:
\begin{equation}
\left| \Psi_0 \right> =  \left| \psi_0 \right> \otimes \left| \chi_0 \right>.
\end{equation}
$\left| \psi_0 \right>$ is the initial state of the central system, for which we chose the singlet state
\begin{equation}
\left| \psi_0 \right>= \dfrac{\left( \left| \uparrow \downarrow \right> - \left| \downarrow \uparrow \right>
\right)}{\sqrt{2}}.
\end{equation}
$\left| \chi_0 \right>$ is the initial state of the bath, for which we chose a statistical superposition of all
basis states with random coefficients.

Depending on the size of the bath we can compute the time evolution either by means of a numerical diagonalization
of the whole Hamiltonian $H$, or by a Chebyshev's polynomial expansion of the time evolution operator
\cite{Dobrovitski:03}. In the former case we can treat up to  $N=11$ spins in the bath finding the exact evolution
of the system. For larger systems we will resort to the approximation of the time evolution operator in a
controlled way. In this section we will treat up to $N=11$ spins, which is large enough for our purposes. In the
next section we will enlarge the bath up to $N=15$ spins using the Chebyshev expansion method.

To obtain a quantitative measure of decoherence, we calculate the reduced density matrix of the system
\begin{equation}
\rho (t) = \text{Tr}_B \left| \Psi (t) \right> \left< \Psi(t) \right|,
\end{equation}
where the subindex $B$ indicates a trace over all degrees of freedom of the bath. In particular, we will analyze
the diagonal elements of the density matrix and the non-diagonal term $\left< \uparrow \downarrow \right| \rho(t)
\left| \downarrow\uparrow \right>$. At $t=0$, $\left< \uparrow \downarrow \right| \rho(t) \left|
\uparrow\downarrow \right>=\left< \downarrow\uparrow \right| \rho(t) \left| \uparrow\downarrow \right>=1/2$,
$\left< \uparrow\uparrow  \right| \rho(t) \left| \uparrow\uparrow \right>=\left< \downarrow\downarrow \right|
\rho(t) \left| \downarrow\downarrow \right>=0$, and $\left< \uparrow \downarrow \right| \rho(t) \left|
\downarrow\uparrow \right>=\left<\downarrow\uparrow \right| \rho(t) \left| \uparrow\downarrow \right>=-1/2$. The
time evolution  shows how the entanglement between the system and the bath destroys the initial correlations of
the system.

Another useful quantity to measure the decoherence is the lineal entropy $\Omega = \text{Tr} \rho^2$. The initial
state of the system is a pure state and the density matrix is idempotent ($\rho^2 = \rho$), thus $\Omega=1$. The
decoherence induced by the bath transforms state of system into mixed state with $\Omega (t) < 1$.  Lower values
of $\Omega$ imply greater efficiency of the decoherence process.

\subsection{Long time evolution of system density matrix}

\begin{figure}[!]
\begin{center}
\rotatebox{-90}{\scalebox{0.28}[0.3]{\includegraphics{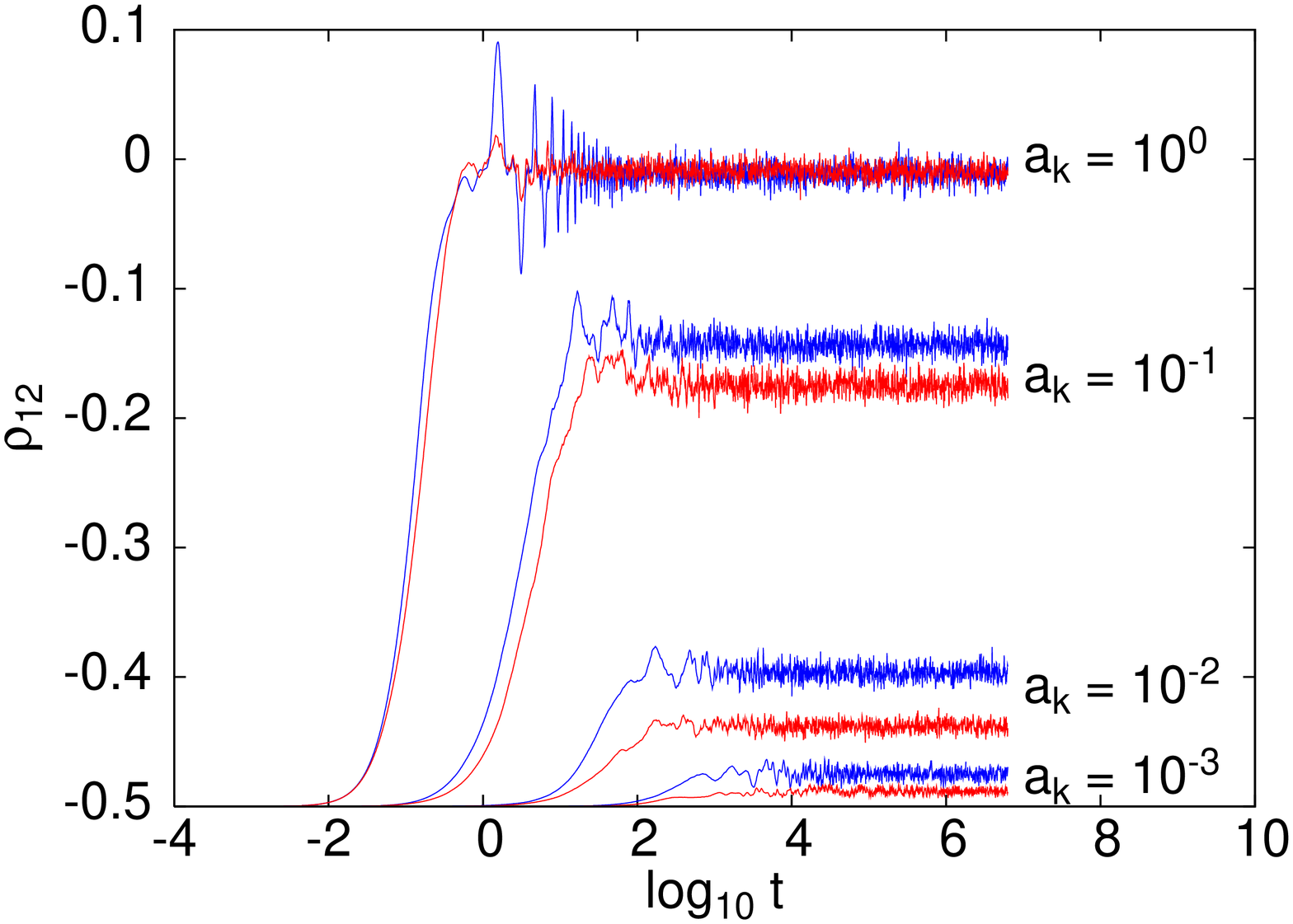}}} \\
\rotatebox{-90}{\scalebox{0.28}[0.3]{\includegraphics{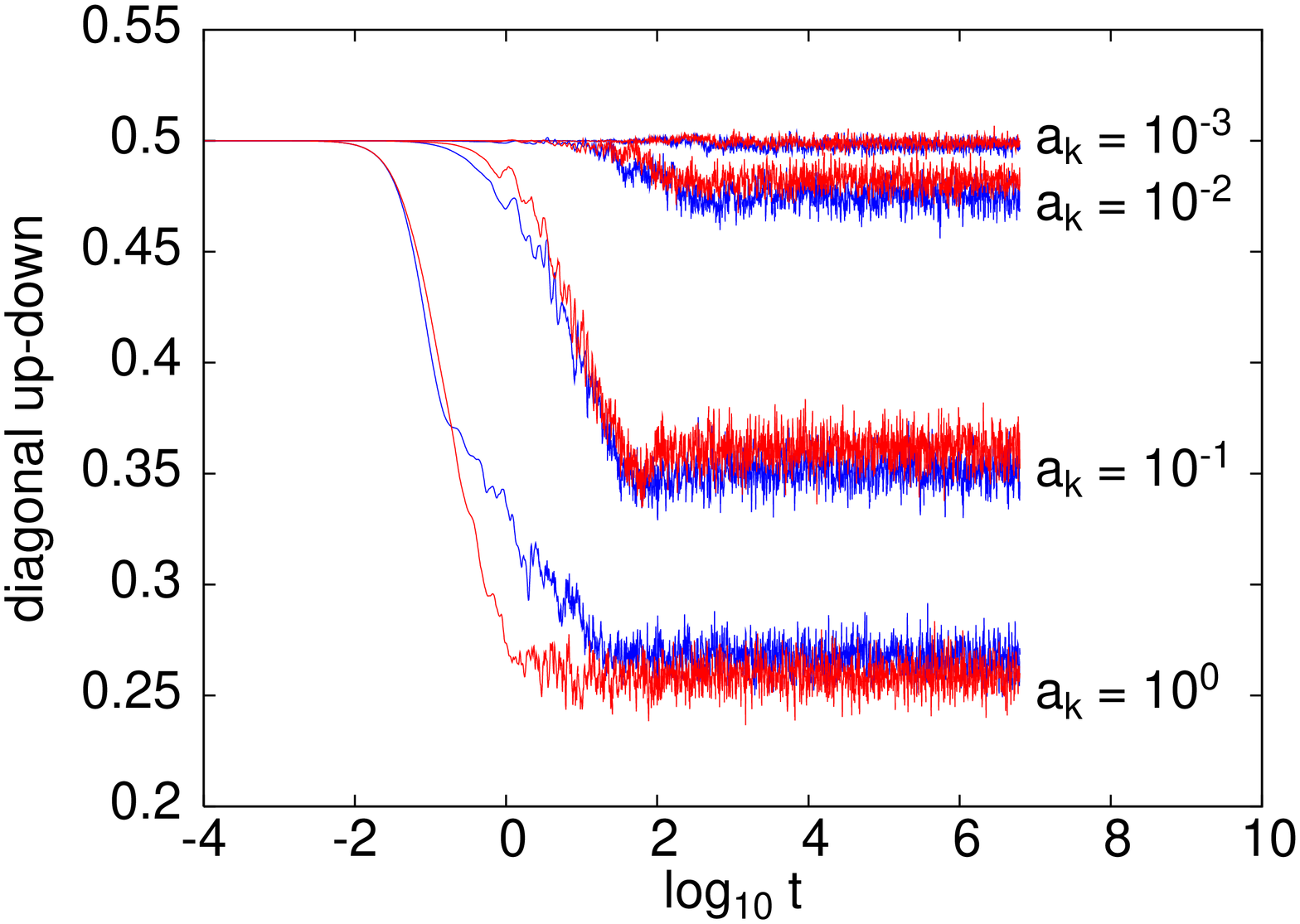}}} \\
\rotatebox{-90}{\scalebox{0.28}[0.3]{\includegraphics{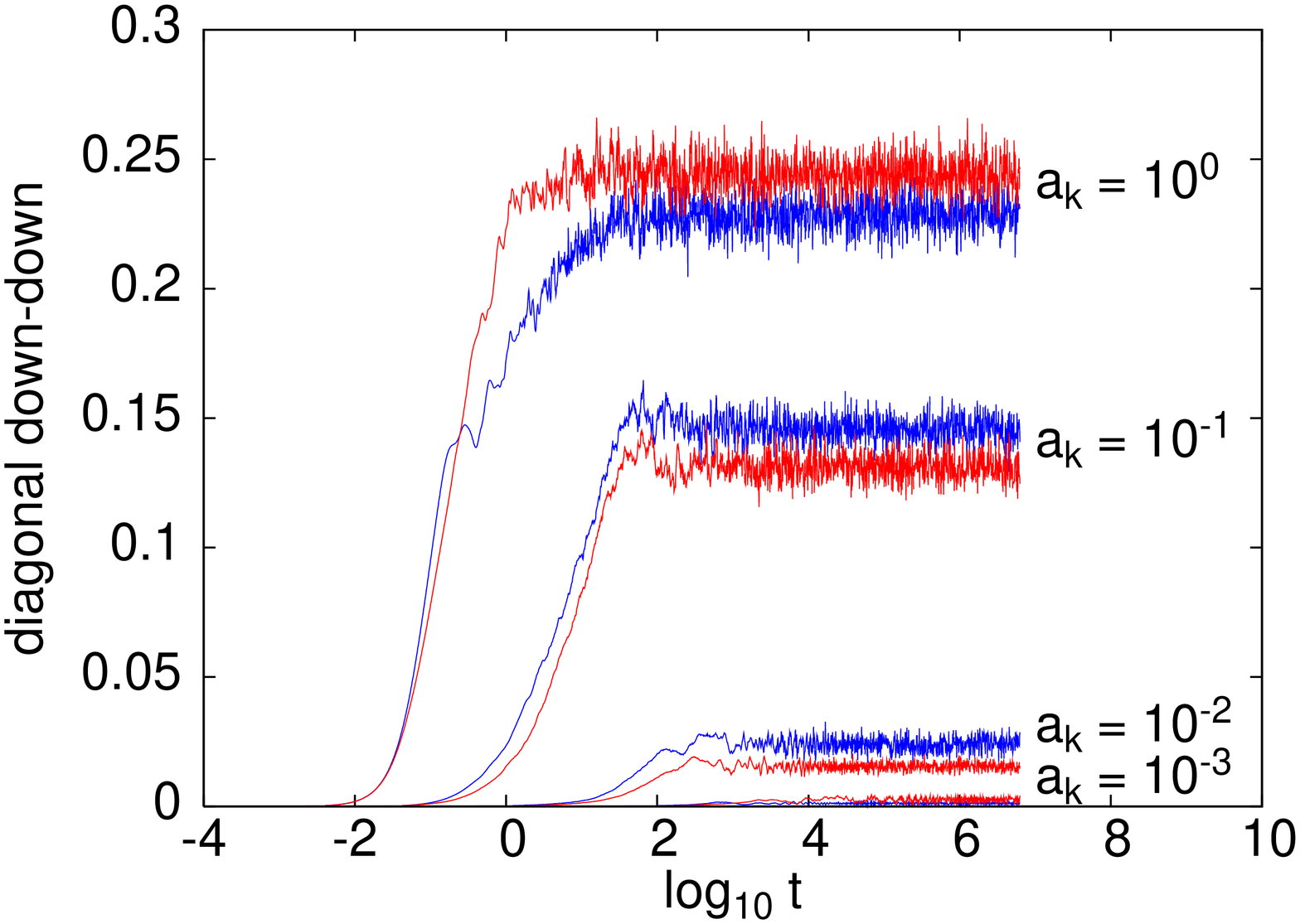}}} \\
\caption[]{Time evolution for different elements of the system density matrix: $\left< \uparrow\downarrow \right| \rho \left| \downarrow\uparrow \right>$ (upper panel), $\left< \uparrow\downarrow \right| \rho \left| \uparrow\downarrow \right>$ (middle panel), and $ \left< \downarrow\downarrow \right| \rho \left| \downarrow\downarrow \right>$ (lower panel), for regular $\alpha=0$ (black line; blue online) and chaotic $\alpha=\pi/2$ (grey line; red online)
limits, as a function of the system-bath coupling strength $a$.} \label{fig:evol_density}
\end{center}
\end{figure}

In Fig. \ref{fig:evol_density} we show the time evolution for $\rho_{12} \equiv \left< \uparrow\downarrow \right|
\rho \left| \downarrow\uparrow \right>$, $\left< \uparrow\downarrow \right| \rho \left| \uparrow\downarrow
\right>$ and $ \left< \downarrow\downarrow \right| \rho \left| \downarrow\downarrow \right>$, as a function of the
parameter $\alpha$ and the system-bath coupling strength $a$, for very long times (note that the time axis is
displayed in logarithmic scale). In all the cases we have chosen $a_k=a \; \forall k$, and $N=11$ spins for the
bath. We can see three main interesting features. First of all, for very long times the elements of the density
matrix of the system relax to equilibrium states, called {\it pointer states}, which are relatively unaffected by
the interaction with the environment and thus survive to the decoherence process. Second, as  expected, the larger
the value of the coupling strength constant the greater the efficiency of the decoherence process. And third, for
small values of $a$, the regular system seems to give rise to a more efficient decoherence than the chaotic one,
whereas for $a \approx 1$, this behavior is reversed.

\begin{figure}[!]
\begin{center}
\rotatebox{-90}{\scalebox{0.28}[0.3]{\includegraphics{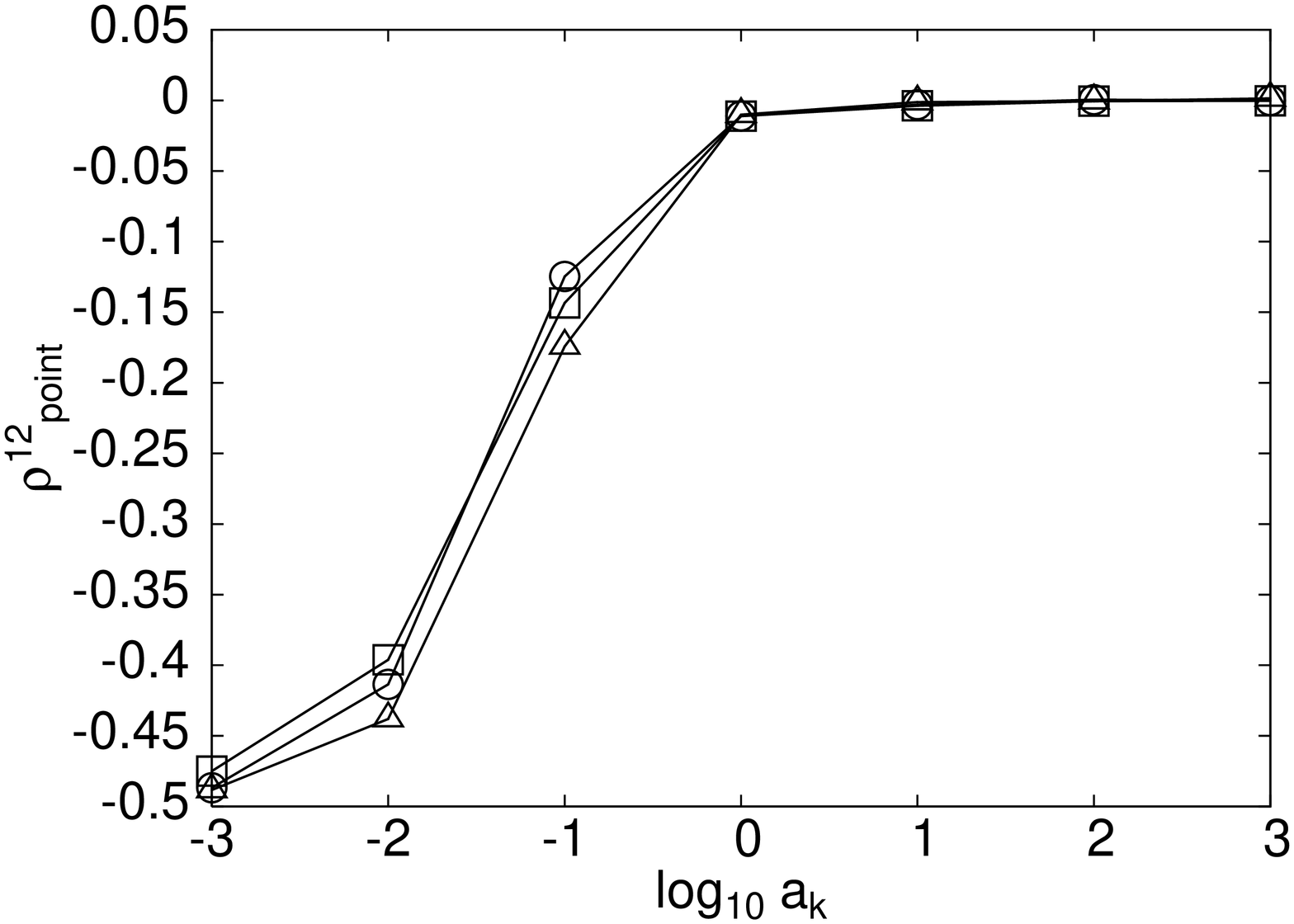}}} \\
\rotatebox{-90}{\scalebox{0.28}[0.3]{\includegraphics{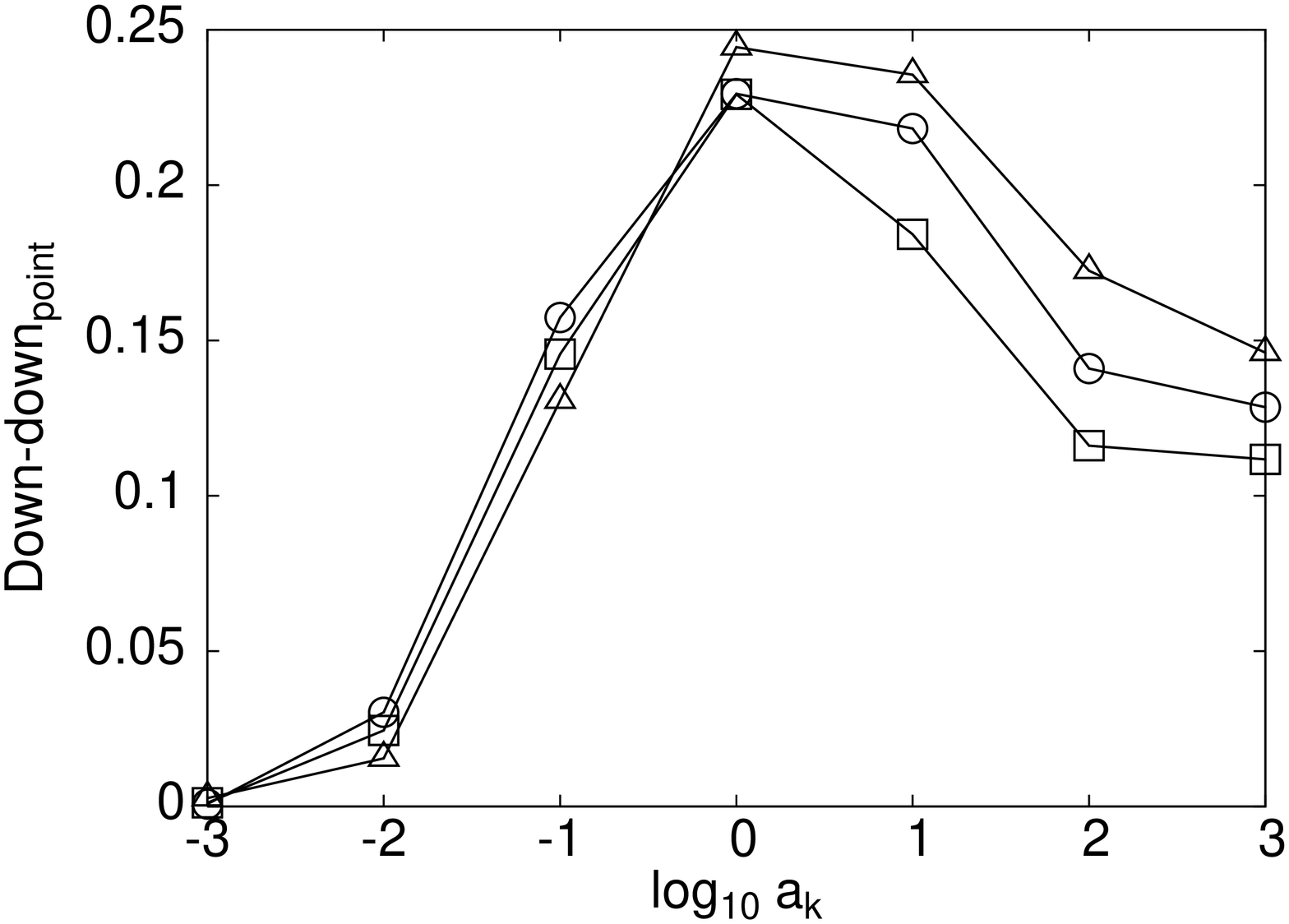}}} \\
\rotatebox{-90}{\scalebox{0.28}[0.3]{\includegraphics{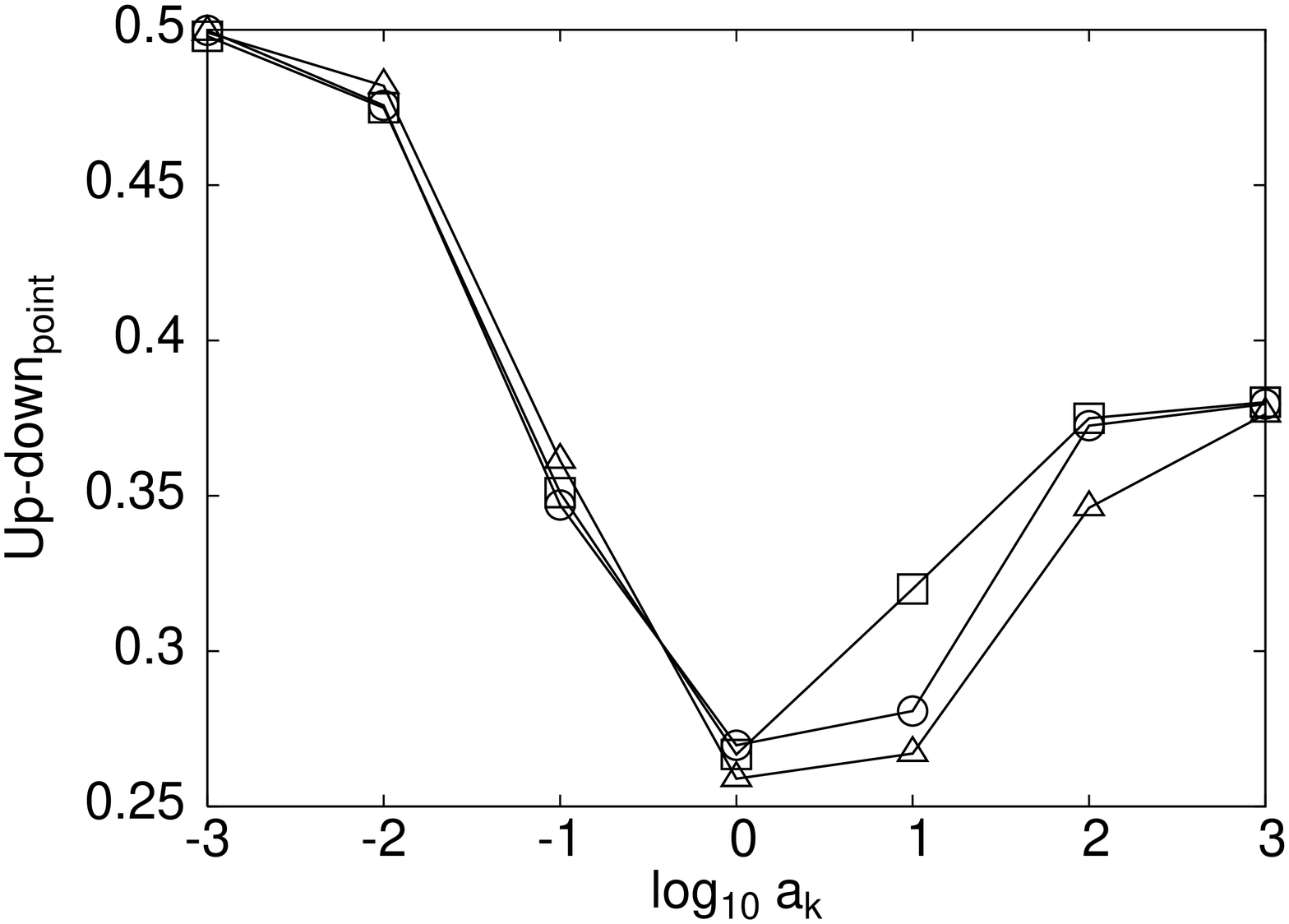}}} \\
\caption[]{Matrix elements of pointer states of the system density matrix: $\left< \uparrow\downarrow \right| \rho \left| \downarrow\uparrow \right>$ (upper panel), $\left< \uparrow\downarrow \right| \rho \left| \uparrow\downarrow \right>$ (middle panel), and $ \left< \downarrow\downarrow \right| \rho \left| \downarrow\downarrow \right>$ (lower panel), for regular $\alpha=0$ (squares) and chaotic $\alpha=\pi/2$ (triangles)
limits, and for an intermediate value $\alpha=\pi/10$ (circles), as a function of the system-bath coupling strength $a$.}
\label{fig:pointers}
\end{center}
\end{figure}

In order to better understand how chaoticity determines the efficency of the bath we calculate the pointer states
elements of the system density-matrix of the previous calculation. The results are shown in Fig.
\ref{fig:pointers}. The elements of the system density-matrix in the pointer states, were obtained by  averaging
the results for $t>10^6$, {\it i. e.} $\rho_{point} = \left< \rho(t) \right>_{t>10^6}$. We can see that in the
three cases the regular limit gives rise to a more efficent decoherence for shorter values of the system-bath
coupling strength, whereas this behavior is reversed for greater values of $a$. In the case of the non-diagonal
$\rho_{12}$ elements, $a \approx 1$ is enough to totally destroy the initial correlations between $\left|
\uparrow\downarrow \right>$ and $\left| \downarrow\uparrow \right>$, characteristics of the initial state
$\psi_0$. For the diagonal elements, however, $a \approx 1$ seems to produce the most efficent decoherence, since
for $a > 1$ the values of the system density-matrix in the pointer state come back to the initial value. In this
last two cases, the change in the relation between decoherence and chaos is clearly seen: for $a < 1$, the regular
limit gives rise to stronger decoherence, while for $a >1$ the chaotic limit becomes a more efficient.

\begin{figure}[!]
\begin{center}
\rotatebox{-90}{\scalebox{0.28}[0.3]{\includegraphics{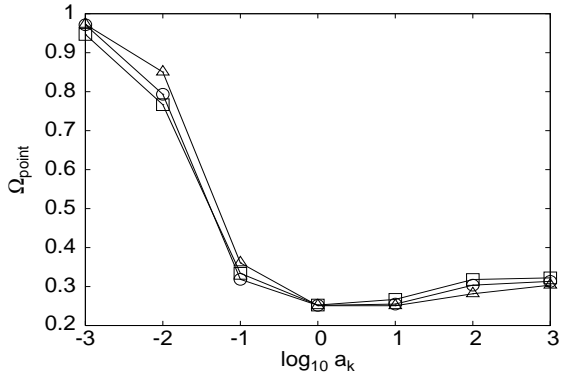}}} \\
\caption[]{Pointer states values of the system linear entropy, for regular $\alpha=0$ (squares) and chaotic
$\alpha=\pi/2$ (triangles) limits, and for an intermediate value $\alpha=\pi/10$ (squares), as a  function of the system-bath coupling
strength $a$.} \label{fig:pointers_omega}
\end{center}
\end{figure}

Similar results are obtained for the linear entropy $\Omega$. Fig. \ref{fig:pointers_omega} shows the pointer
state values of $\Omega$ for $\alpha=0$, $\alpha=\pi/10$ and $\alpha=\pi/2$. It is clearly seen that for $a < 1$,
the regular bath produces a stronger decoherence, whereas for $a > 1$ this behavior is reversed. We can conclude
that the dynamical regime of the bath determines the efficiency of the decoherence process. Moreover, the
system-bath coupling strength determines whether integrability or chaos give rise to more efficient decoherence.

\section{Decoherence time and short time evolution of system density matrix in perturbative regime}

Having established the relation between the efficiency of decoherence and the dynamical regime of the bath for
different values of the system-bath coupling strength, we now focus on the quantitative analysis of how quantum
chaos affects the decoherence process. Our aim is to determine whether a smooth transition from integrability to
chaos in the bath can be detected from quantities directly related to decoherence. For this purpose, we select a
small value for the system-bath coupling strength, for which the integrable limit produces a stronger decoherence
than the chaotic one. This choice allows us to follow the central system decoherence by its fidelity $F(t)$,
analogous to the Loschmidt echo, which measures the sensitivity of the system to external perturbations. The
fidelity is defined as
\begin{equation}
F(t) = \text{Tr}_S \left[ \rho'(t) \rho(t) \right],
\end{equation}
where $\rho'(t)$ is the system density matrix for an ideal evolution in which the system and the bath do not
interact, {\it i. e.} $H_{SB}=0$. The subindex $S$ denotes a trace over the states of the central system. This
quantity behaves in a similar way as the linear entropy $\Omega$.

In what follows, we will consider a bath composed of $N=15$ spins. The evolution is approximated by means a
Chebyshev expansion of the evolution operator. We have checked that the size of the bath does not change
qualitatively the decoherence process, but the fluctuations around the pointer states (see Fig
\ref{fig:evol_density}) are decreased.  The system-bath coupling strength is set to $a_k = \sqrt{\frac{15}{11}}
10^{-2}$ $\forall k$. This particular value was chosen because  the coupling behaves as $b=\sqrt{\sum_{k=1}^N
a_k^2}$ \cite{Lages:05}, therefore the set of $\{ a_k \}$ has to be scaled with the bath size.

\begin{figure}[!]
\begin{center}
\rotatebox{-90}{\scalebox{0.28}[0.3]{\includegraphics{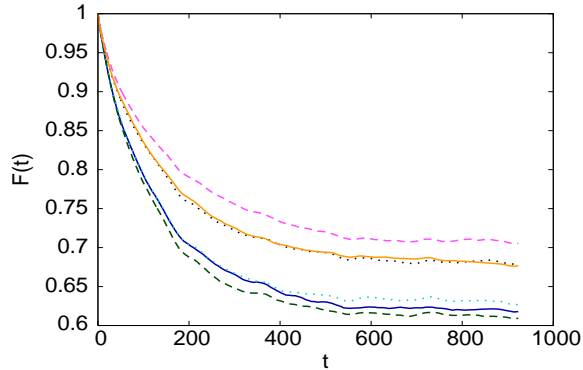}}} \\
\caption[]{Fidelity $F(t)$ for the central system perturbed by a bath composed by $N=15$ spins with $a_k = \sqrt{\frac{15}{11}} 10^{-2} ~ \forall k$, for different values of parameter $\alpha$: solid black (blue online) corresponds to $\alpha=0$; dashed black (green online) corresponds to $\alpha=\pi/20$; dotted grey (cyan online) corresponds to $\alpha=\pi/10$; dotted black (black online) corresponds to $\alpha=\pi/5$; dashed grey (magenta online) corresponds to $\alpha=3 \pi /10$; and solid grey (orange online) corresponds to $\alpha=\pi/2$.}
\label{fig:eco}
\end{center}
\end{figure}

In Fig. \ref{fig:eco} we show the fidelity $F(t)$ for different values of parameter $\alpha$. Several interesting
facts emerge from the figure. First of all, the shape of all the curves is very similar. The main differences
between them are related to the pointer states, {\it i. e.} the values reached after a long time evolution. We can
also see that these pointer values of the fidelity do not increase monotonically with parameter $\alpha$, as it is
expected since the integrable limit gives rise to a stronger decoherence than the chaotic one for this value of
the system-bath coupling strength. The numerical results show that when $\alpha$ is close to zero and, thus, the
bath is close to integrability, the decoherence is more efficient than when $\alpha$ is close to $\pi/2$, that is,
when the bath is close to the fully chaotic limit. However, this efficiency does not decrease monotonically for
increasing $\alpha$. In fact, the curves in the figure show that for $\alpha=\pi/20$, the fidelity is smaller than
for $\alpha=0$, and for $\alpha=3 \pi /10$, it is larger than for $\alpha=\pi/2$. Secondly, the transition from
the values characterizing integrability to those corresponding to chaos is slower than the transition from
integrability to chaos determined by the spectral statistics. For example, for $\alpha=\pi/10$, the fidelity is
close to the integrable limit $\alpha=0$, whereas the spectral statistics are closer to the chaotic limit. In
particular, the parameter $\eta$ indicates an almost chaotic behavior for $\alpha=\pi / 10$ (note that from the
results shown in Fig. \ref{fig:eta} we can conclude that an increase of the size of the bath accelerates this
transition, and therefore we may expect that for $N=15$ and $\alpha \approx \pi / 20$ the bath is almost chaotic).
The power spectrum of $\delta_n$ statistic is also close to the theoretical value for the fully chaotic limit, but
it reveals that such limit is not yet reached.

Another important quantity related to the decoherence process is the decoherence time, that is, the characteristic
time for the loss of coherence of the central system due to the coupling with the bath. One way to estimate this
time is by means of the decay that the bath induces in $F(t)$. We can fit the shape of $F(t)$ to the following
expression
\begin{equation}
F(t)=p + (1-p) \exp \left(-t^{\beta}/T_d^{\beta} \right), \label{eq:fidelity}
\end{equation}
where $p$ is the pointer value for $F(t)$, and $\beta$ and $T_d$ are free parameters, the last one corresponding
to the decay time of the system. The results for this fit are shown in table \ref{tab:eco}. We can see that the
decoherence time is slightly larger for a chaotic bath, and that there is a clear correlation between the pointer
value $p$ and the decay time $T_d$. Moreover, table \ref{tab:eco} also displays a surprising result: contrary to
what is expected, $F(t)$ decays roughly in an exponential way for all the values of $\alpha$. Therefore, it seems
that the dynamical regime of the bath does not affect the decay ratio of the Fidelity.

\begin{table}
\begin{tabular}{c||c|c|c}
$\alpha$ & $p$ & $\beta$ & $T_s$ \\ \hline \hline
$0$ & $0.6200$ & $0.8781$ & $127.1$  \\
$\pi/20$ & $0.6118$ & $0.9146$ & $121.0$ \\
$\pi/10$ & $0.6306$ & $0.9036$ & $121.0$ \\
$\pi/5$ & $0.6817$ & $0.8691$ & $134.9$ \\
$3 \pi / 10$ & $0.7088$ & $0.8402$ & $148.2$ \\
$\pi / 2$ & $0.6797$ & $0.8589$ & $135.0$ \\ \hline
\end{tabular}
\caption[]{Parameters of eq. (\ref{eq:fidelity}) for different values of $\alpha$.} \label{tab:eco}
\end{table}

\begin{figure}[!]
\begin{center}
\rotatebox{-90}{\scalebox{0.28}[0.3]{\includegraphics{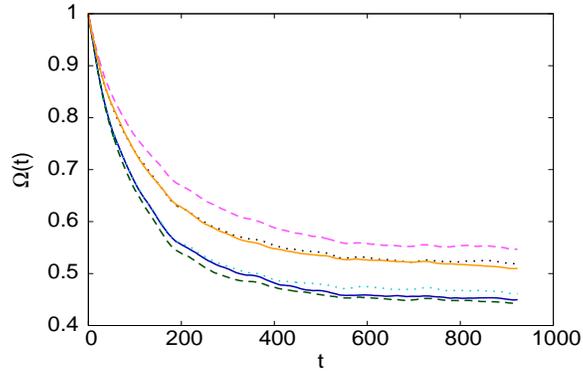}}} \\
\caption[]{Linear entropy $\Omega(t)$ for the central system perturbed by a bath composed by $N=15$ spins with $a_k = \sqrt{\frac{15}{11}} 10^{-2} ~ \forall k$, for different
values of parameter $\alpha$: solid black (blue online) corresponds to $\alpha=0$; dashed black (green online) corresponds to $\alpha=\pi/20$; dotted grey (cyan online) corresponds to $\alpha=\pi/10$; dotted black (black online) corresponds to $\alpha=\pi/5$; dashed grey (magenta online) corresponds to $\alpha=3 \pi /10$; and solid grey (orange online) corresponds to $\alpha=\pi/2$.} 
\label{fig:entropia}
\end{center}
\end{figure}

The close connection between the fidelity $F(t)$ and the linear entropy $\Omega(t)$ can be seen in Fig.
\ref{fig:entropia}. The shape of the curves for linear entropy and fidelity are almost identical. The only
appreciable difference between these two quantities is their pointer value. Moreover, the transition from
integrability to chaos follows the same trend: it is non-monotonic with $\alpha$, and slower than the
corresponding transition in the spectral statistics.

A characteristic time of decoherence can be also defined from $\Omega(t)$ fitting the numerical results to an
expression similar to \eqref{eq:fidelity}. We show the results in table \ref{tab:time}. They are very similar to
those obtained with the fidelity: for values of $\alpha$ closer to the integrable limit, decoherence takes place
faster than for values closer to the fully chaotic regime. There is also a strong correlation between the pointer
value and the decoherence time. We also remark that the transition is not monotonous with $\alpha$.

\begin{table}
\begin{tabular}{c||c|c|c}
$\alpha$ & $p$ & $\beta$ & $T_s$ \\ \hline \hline
$0$ & $0.4522$ & $0.8801$ & $112.5$  \\
$\pi/20$ & $0.4459$ & $0.9078$ & $106.5$ \\
$\pi/10$ & $0.4655$ & $0.8994$ & $107.7$ \\
$\pi/5$ & $0.5228$ & $0.8796$ & $125.0$ \\
$3 \pi / 10$ & $0.5515$ & $0.8599$ & $140.6$ \\
$\pi / 2$ & $0.5138$ & $0.8700$ & $130.2$ \\ \hline
\end{tabular}
\caption[]{Parameters of eq. (\ref{eq:fidelity}) applied to lineal entropy $\Omega(t)$ for different values of
$\alpha$.} \label{tab:time}
\end{table}

From all these results we can conclude that the dynamical regime of the bath influences the pointers states of the
central system in a non-strictly monotonic way. The modifications of the pointer states are not smooth nor
monotonic when the bath change from integrability to chaos. We have also shown that the decoherence time is
influenced by the dynamical regime of the bath in a very similar way: from the decay of both Loschmidt echo and
linear entropy, a trend from integrability to chaos can be identified in the decoherence time. However, it is also
important to note that the rate of this decay is not affected by the dynamical regime of the bath, since it is
roughly exponential along the whole transition. A possible explanation of this surprising feature is that the
complexity of the Hamiltonian remains more or less the same along the whole transition from integrability to
chaos, because it always consists on a XYZ model in which none of the terms is negligible.

\section{Conclusions}

We have studied the decoherence of a two-spin central system interacting with a bath whose dynamical regime can
transit from integrability to chaos. Unlike previous studies the integrable regime is described by an  XYZ Gaudin
magnet with random parameters, with a complexity similar as the corresponding to the fully chaotic regime. We have
calculated time evolution by numerical technics, and we have analyzed several quantities related to the reduced
density matrix of the central system $\rho(t)$.

From our results, we conclude that at $t \longrightarrow \infty$ the reduced density matrix of the system $\rho$
has an observable dependence on the dynamical regime of the bath. For small values of the system-bath coupling
strength, the asymptotic value of $\left< \uparrow \downarrow \right| \rho \left| \downarrow \uparrow \right>$ is
larger in the regular limit than in the chaotic limit. However, this difference tends to decrease as the
system-bath coupling strength is increased. For the diagonal elements $\left< \uparrow \downarrow \right| \rho
\left| \uparrow \downarrow \right>$ and  $\left< \downarrow \downarrow \right| \rho \left| \downarrow \downarrow
\right>$, the relation between the dynamical regime of the bath and the pointer states changes at $a \approx 1$.
 Below this value, the regular limit gives rise to a stronger decoherence; above it, the efficiency of the decoherence
process is larger when the bath is chaotic. For both diagonal elements $a \approx 1$ gives rise to the larger
decoherence. These results show that the onset of chaos affects the decoherence process of the central system in a
non simple way, since its influence depends on how strong is the coupling with the bath. These conclusions are
consistent with those obtained in \cite{Lages:05}.

The connection between the properties of the reduced density matrix of the system and the dynamical regime of the
bath allows to analyze the transition form integrability to chaos in terms of decoherence. We have done so with a
small value of the system-bath coupling-strength, that is, in the perturbative regime. Our results show that this
transition is not so smooth and monotonous as described by the spectral statistics analysis, in spite of the fact
that the regular and chaotic limits are clearly distinguished. Beyond this non-monotonic behavior, we have also
shown that the transition from integrability to chaos in the central system reduced density matrix is slower that
than the transition in the nearest neighbor spacing distribution by means of $\eta$ parameter, and it is a bit
closer to the behavior of the power spectrum of $\delta_n$ statistic. Therefore, long-range correlations in
spectral fluctuations seem to be involved in the bath efficiency to produce decoherence.

We have also performed a similar analysis with the decoherence time, that is, the time at which the central system
losses its original correlations. For a perturbative regime, the integrable limit produces stronger decoherence in
shorter times as compared with the fully chaotic regime. Nevertheless, in contradiction with the results of
reference \cite{Lages:05}, the decay rate of the fidelity and the linear entropy does not depend on the dynamical
regime of the bath. The main difference between both treatments is the modelling of the integrable regime; while
our integrable limit contains all the complexity of the chaotic regime, reference \cite{Lages:05} uses a simple
integrable limit with large degeneracies. Therefore, we conclude that the decay of the fidelity and the linear
entropy is related to the complexity of the bath and not to its dynamical regime. Further work is needed to
clarify this result.

\section*{Aknowledgments}

This work was supported by grants FIS2006-12783-C03-01 from Ministerio de Educaci\'on y Ciencia of Spain, and
200650M012 from Comunidad de Madrid and CSIC. A.R. is supported by the Spanish program "Juan de la Cierva". R.A.M.
is supported by the I3P program funded by the European Social Fund.


\begin{thebibliography}{999}

\bibitem{Zurek:03} W. H. Zurek, Rev. Mod. Phys. {\bf 75}, 715 (2003).

\bibitem{Nielsen} M. Nielsen and I. Chuang, {\it Quantum Computation and Quantum Information} (Cambridge University Press, Cambridge, UK, 2000).

\bibitem{Cucchietti:03} F. M. Cucchietti, D. A. R. Dalvit, J. P. Paz, and W. H. Zurek, Phys. Rev. Lett. {\bf 91}, 210403 (2003).

\bibitem{lyapunov} W. H. Zurek and J. P. Paz, Phys. Rev. Lett. {\bf 72}, 2508 (1994); P. A. Miller and S. Sarkar, Phys. Rev. E {\bf 58}, 4217 (1998); {\it ibid} {\bf 60}, 1542 (1999); A. K. Pattanayak, Phys. Rev. Lett. {\bf 83}, 4526 (1999); D. Monteoliva and J. P. Paz, Phys. Rev. Lett. {\bf 85}, 3373; {\it ibid} Phys. Rev. E {\bf 64} 056238 (2001); R. A. Jalabert and H. M. Pastawski
Phys. Rev. Let. {\bf 86} 2490 (2001); P. Bianucci, J. P. Paz, and M. Saraceno, Phys. Rev. E {\bf 65} 046226 (2002).

\bibitem{Zurek:01} W. Zurek, Nature (London) {\bf 412}, 712 (2001).

\bibitem{Dobrovitski:03} V. V. Dobrovitski and H. A. De Raedt. Phys. Rev. E {\bf 67}, 056702 (2003).

\bibitem{Blume:03} R. Blume-Kohout and W. Zurek, Phys. Rev. A {\bf 68}, 032104 (2003); F. C. Lombardo and P. I. Villar, Phys. Rev. A {\bf 72} 034103 (2005).

\bibitem{Paz:06} L. Ermann, J. P. Paz, and M. Saraceno, Phys. Rev. A {\bf 73}, 012302 (2006).

\bibitem{Hou:04} X.-W. Hou and B. Hu, Phys. Rev. A {\bf 69} 042110 (2004).

\bibitem{regular} A. Tanaka, J. Phys. A {\bf 29}, 5475 (1996); R. M. Angelo, K. Furuya, M. C. Nemes, and G. Q. Pellegrino, Phys. Rev. E {\bf 60}, 5407 (1999).

\bibitem{regular_decoherencia} T. Prosen and M. Znidaric, J. Phys. A {\bf 35}, 1455 (2002).

\bibitem{Lages:05} J. Lages, V. V. Dobrovitski, M. I. Katsnelson, H. A. De Raedt, and B. N. Harmon, Phys. Rev. E {\bf 72}, 026225 (2005).

\bibitem{Gould:02} M. K. Gould, Y.-Z. Zhang, and S.-Y. Zhao, Nucl. Phys. B, 492 (2002).

\bibitem{nota1} We use different notation for the central spins, {\bf S$_i$}, and for bath spins, {\bf I$_k$}, just to distinguish them more easily. Both central and bath are identical 1/2 spins. 

\bibitem{Katsnelson:03} M. I. Katsnelson, V. V. Dobrovitski, H. A. De Raedt, and B. N. Harmon, Phys. Lett. A {\bf 318}, 445 (2003).

\bibitem{nota2} We are implicitly considering the following definition of integrability: a quantum system is said to be integrable if a set of as many conmuting Hermitian operators as quantum degrees of freedom can be explicitly given, and the Hamiltonian can be expressed as a function of these operators \cite{Weigert:95}.

\bibitem{Weigert:95} S. Weigert and G. Muller, Chaos, Solitons and Fractals, {\bf 5}, 1419 (1995).

\bibitem{Relano:04} A. Rela\~no, J. Dukelsky, J. M. G. G\'omez, and J. Retamosa, Phys. Rev. E {\bf 70}, 026208 (2004).

\bibitem{Berry:77} M. V. Berry and M. Tabor, Proc. R. Soc. London, Ser. A {\bf 356}, 375 (1977).

\bibitem{Bohigas:84} O. Bohigas, M. J. Giannoni, and C. Schmit, Phys. Rev. Lett. {\bf 52}, 1 (1984).

\bibitem{Haake} F. Haake, {\it Quantum Signatures of Chaos} (Springer-Verlag, Berlin, 2001).

\bibitem{delta_n} A. Rela\~no, J. M. G. G\'omez, R. A. Molina, J. Retamosa, and E. Faleiro, Phys. Rev. Lett. {\bf 89}, 244102 (2002); E. Faleiro, J. M. G. G\'omez, R. A. Molina, L. Mu\~noz, A. Rela\~no, and J. Retamosa, Phys. Rev. Lett. {\bf 93} 244101 (2004).

\bibitem{Gomez:05} J. M. G. G\'omez, A. Rela\~no, J. Retamosa, E. Faleiro, L. Salasnich, M. Vranicar, and M. Robnik, Phys. Rev. Lett. {\bf 94} 084101 (2005).

\bibitem{Gomez:02} J. M. G. G\'omez, R. A. Molina, A. Rela\~no, and J. Retamosa, Phys. Rev. E {\bf 66}, 036209 (2002).

\bibitem{Marquardt:06} F. Marquardt in {\em Advances in Solid State Physics},
Vol. 46, ed. R. Haug, Springer 2006.



\end{thebibliography}
\end{document}